\documentclass[aps,amsmath,amssymb,showkeys]{revtex4}
\usepackage{graphicx}
\usepackage{dcolumn}
\usepackage{bm}

\newcommand{\be}{\begin{equation}}
\newcommand{\ee}{\end{equation}}
\newcommand{\bea}{\begin{eqnarray}}
\newcommand{\eea}{\end{eqnarray}}
\newcommand{\nn}{\nonumber\\}
\def\journal#1#2#3#4{{#1} {\bf #2}, #3 (#4)}

\def\la{\langle}
\def\ra{\rangle}
\def\eq#1{(\ref{#1})}
\def\hf{\frac{1}{2}}
\def\ve#1{\mathbf{#1}}
\def\mr#1{\mathrm{#1}}
\def\ord#1{{\cal O}\left(#1\right)}

\def\tr{\mr{tr}}

\def\bphi{\bar\phi}
\def\bphid{\bar\phi^\dagger}
\def\chid{\chi^\dagger}
\def\phid{\phi^\dagger}

\begin{document}
\title{Yang-Mills-Higgs models with higher order derivatives}
\author{Janos Polonyi$^a$, Alicja Siwek$^{b}$}
\affiliation{$^a$University of Strasbourg,
High Energy Physics Theory Group, CNRS-IPHC,
23 rue du Loess, BP28 67037 Strasbourg Cedex 2, France}
\affiliation{$^b$Wroc{\l}aw University of Technology, Institute of Physics, Wybrze\.ze~Wyspia\'nskiego~27, 50-370 Wroc{\l}aw, Poland}
\begin{abstract}
Models with higher order derivative terms in the kinetic energy appear not only as effective theories, they can be considered as elementary, renormalizable models in their own right. The extension of Higgs mechanism is discussed for Yang-Mills-Higgs models where the kinetic energy of the scalar field contains higher order derivatives. The connection term of the covariant derivatives generates a local potential for the gauge field in the phase with spontaneously broken global gauge symmetry which may break Lorentz symmetry in the vacuum. Relativistic, massless gauge bosons spectrum is found for small fluctuations where the only remnant of the breakdown of the Lorentz symmetry is a shift of the energy induced by the expectation value of the temporal component of the gauge field, reminiscent of a chemical potential.
\end{abstract}
\date{\today}
\maketitle

\section{Introduction}
There are two different ways in which higher order derivative terms may arise in the Lagrangian of field theoretical models. The usual way is to start with an elementary, renormalizable theory and look for the effective dynamics, below some heavy mass particle production threshold. The effective Lagrangian usually contains space-time derivatives of arbitrary orders, reflecting the nonlocality of the dynamics at the characteristic scale of the heavy particle. Another possibility is to include some higher order derivatives in the Lagrangian of an elementary, renormalizable model as a part of the dynamics or simply as a regulator \cite{dirac,podolski,pauli,schwed,villars,feynman,leewick}. The development of both kinds of models is seriously hindered by notorious problems with unitarity \cite{pais}. Though this is a trivial problem in the case of the effective theory, arising from the incompatibility of a smooth UV cutoff in energy and the truncation of the gradient expansion, there is no satisfactory solution in sight \cite{boulware}. 

The imposition of the most natural condition to ensure unitarity, namely to keep the elementary excitation spectrum real, leads to an unexpected difficulty: theories with higher order derivatives handle states with positive, negative and vanishing norm. States with negative norm defy physical interpretation, the weight of states with vanishing norm remains unbounded even for unitary time dependence and may generate instabilities. But the unitarity of the time evolution can be confirmed in PT-invariant theories because the signature of the norm is given by the PT parity \cite{cons,dyn}.

Once the troubling aspects of the higher order derivative terms can be controlled it is natural to inquire if such theories can display new type of phenomenology. There are two features of theories with higher order derivatives which make them unique. One of them is a very economical unification. It is well known in solid state physics that higher order time derivatives in the kinetic energy lead to several bands. In the present context of relativistic models the higher order space-time derivatives in the kinetic term generate several relativistic particles, preons. Two nontrivial manifestations of preons in the presence of spontaneously broken continuous symmetry are considered in this work. It is shown below that the spontaneous breakdown of a global, continuous symmetry leads to massless Goldstone preons as expected, leaving the possibility for the remaining preons of the Goldstone mode to be massive. In case of a local symmetry the Higgs mechanism is modified. The gauge field remains massless and the usual balance between the disappearing Goldstone modes and the appearing longitudinal modes of the gauge field, observed in the usual Higgs mechanism does not hold anymore. In particular, the number of Goldstone preons of the scalar field which are removed from the physical, gauge invariant sector of the theory is not necessarily related to the number of preons appearing in the massless gauge field. 

Another unusual possibility, offered by the higher order covariant derivative terms in gauge theories is the spontaneous breakdown of Lorentz symmetry. The status of Lorentz symmetry is remarkably nontrivial in particle physics. Einstein's special relativity relies on macroscopic objects like meter rods and watches and is lacking its solid observational basis in the microscopic, quantum regime. The rules of formal implementation of Lorentz symmetry in QED are confirmed by the high precision measurements but their relation to Einstein's special relativity is less clear in the absence of measuring devices of velocity of propagation within an atom or a nucleon. The retarded propagators, describing the linear response to an external source, are supposed to satisfy causality and are made to vanish for space-like separation. But the Feynman propagator is nonvanishing for space-like separation allowing acausal coupling among quantum fluctuations, a possible source of acausal higher order, nonlinear response. 

This issue can be raised in a more phenomenological manner, as well, by recalling an important lesson of the renormalization group method: the measured values depend on the scale of observation. In other words there are no constants in physics. Any experimental constant is actually a relatively large plateau with slowly varying scale dependence only. Is the velocity of light an exception to this rule? Lorentz symmetry is obviously lost at long distances due to the expansion of the Universe. Since the experimental verification of special relativity is difficult at extremely short distances one is tempted to look for possible quantum effects, generating different speed of light at microscopic scales without unsettling Einstein's special relativity stated and experimentally verified beyond any doubt at intermediate scales. 

Actually, one stumbles on a problem with Lorentz symmetry at short distances independently of the suspicion raised by renormalization group. It is the impossibility of constructing UV regulator with relativistic symmetries. The reason is the impossibility of defining a region in the energy-momentum space in terms of invariant length due to the divergent volume of the Lorentz group. The regulators, based on Wick-rotation always contain an $\ord{1/\Lambda}$ Lorentz violating part, $\Lambda$ being the energy scale of the UV cutoff. This is not a problem for renormalizable non-gravitational theories where the global Lorentz symmetry is recovered when the cutoff is removed \cite{anselmi} but poses a serious problem for quantum gravity where the gauge symmetry, involving Lorentz transformations must be implemented in a cutoff-independent manner. The way out might be either the use of stringy description as a regulator or giving up altogether the Lorentz structure at the cutoff \cite{horava,visser}. The recent observations about the possibility of generalizing special relativity to include the Planck length \cite{camelia,magueiko} represent a promising step in constructing an interpolating theory.

Therefore one turns to the question of the breakdown and the restoration of Lorentz symmetry in the function of the scale of observation in quantum field theory. The simplest way to break a symmetry is to introduce explicit, non-symmetrical terms in the action. But it may happen that a symmetry, present formally on the level of the bare action is broken dynamically at some finite scale. If the symmetry breaking is driven by asymptotically long distance modes then it is called spontaneous. It was found a long time ago \cite{nielsen,chada} that the explicit breaking of Lorentz invariance, introduced by the specific choice of metric tensor or the vierbein at the cutoff scale, is suppressed at finite, physical scales as the cutoff is removed. If the Lorentz symmetry is broken explicitly in a Yukawa model at the cutoff scale by non-derivative couplings then it is found to be restored at finite energies after the cutoff is removed \cite{gomes}. A particular symmetry breaking, characterized by a four-vector in a Chern-Simons action has been proposed in a CPT-even manner \cite{carroll}. Another symmetry breaking pattern, achieved by dimension-5 operators has been identified in Ref. \cite{myers}. The issue of generating Lorentz breaking terms by radiative corrections from simpler theory with CPT-odd and non-Lorentz-invariant terms in its action has been considered as well \cite{calladayk,coleman,jackiw,mariz}.

The astrophysical long distances violation of Lorentz symmetry hints at a spontaneous symmetry breaking, a gravitational Higgs mechanism \cite{kostelecky} and relations between gravitons and Goldstone modes \cite{kraus,kosteleckygm}, a scenario based on an earlier, Abelian idea \cite{heisenberg,bjorken,eguchi}. One can imagine non-gravitational roots of the spontaneous breakdown of Lorentz symmetry. In a generic relativistic model the nonvanishing vacuum expectation value of a vector or tensor field signals such a symmetry breaking \cite{kostelecky,kosteleckybb}. This order parameter is not necessarily an elementary field, it can be generated dynamically \cite{jenkins}, as well. It should be noted that the spontaneous breakdown of Lorentz symmetry might by hidden from observation, a point of view which leads to the systematic establishment of gauge theories \cite{holger}. 

A simple dynamical breakdown of Lorentz symmetry has already been demonstrated in an effective scalar QED \cite{dyn}. When higher order covariant derivatives occur in the kinetic energy of a charged scalar particle then the connection term of the covariant derivative induces a local potential for the gauge field and this may lead to nonvanishing vacuum expectation value for the gauge field. The excitation spectrum is not necessarily relativistically invariant in this phase therefore one has to distinguish carefully the relativistic, massless preon spectrum from the relativistically noninvariant, gapless spectrum. In case of scalar QED the gauge field is the Goldstone mode, arising from the spontaneous breakdown of the symmetry with respect to Lorentz boosts and remains gapless. The usual, relativistic Green functions are recovered for the gauge field, pushing the signature of the breakdown of the Lorentz symmetry beyond the tree-level dynamics of the photon field. We consider a non-Abelian Yang-Mills-Higgs model in this work where the naive argument would allow the mass generation for the gauge field in the generalized Higgs mechanism. Instead, the tree-level dynamics of the gauge field remains gapless and the only breakdown of Lorentz invariance is the emergence of a shift of the energy of the charged gauge field components, reminiscent of a chemical potential, provided by the expectation value of the temporal component of the gauge field.

The local potential, generated by the connection term of the covariant derivatives in the spontaneously broken phase of the global gauge symmetry is reminiscent of the bumblebee models \cite{kosteleckybb}. The Lorentz symmetry is then recovered on the tree \cite{nambu} and one-loop level \cite{azatov,chkareuliqed} and with non-Abelian symmetry \cite{chkareuliym}. This was argued to be a mechanism to render the breakdown of Lorentz symmetry invisible \cite{chkareuliholger}. Our model differs from those used in these considerations because our local potential for the gauge field leaves the invariance under local gauge transformations intact.

The preon content of a scalar field with a global internal symmetry is discussed in Section \ref{preoncs}, starting with the well known difficulties, caused by the higher order derivatives in the kinetic energy in Section \ref{diffs}. The preon fields and condensate are introduced in Section \ref{preonss} and \ref{afvac}, respectively. The extension of Goldstone theorem for preons is presented in the simple context of an effective potential in Section \ref{godstprs}. The extension to local symmetry is addressed in Section \ref{gaguges}, where first the tree-level vacuum is identified in Section \ref{meanfvs}, followed by the calculation of the quadratic action for small fluctuations in Section \ref{flucts}. The gapless nature of the scalar sector is verified in Section \ref{gaplesss}. Section \ref{utgths} contains a more detailed discussion of a simpler case where the gauge group is chosen to be $U(2)=SU(2)\times U(1)$. The gauge field condensate can be either purely non-Abelian or completely Abelian in this case, these possibilities are considered in Sections \ref{nagcs} and \ref{agcs}, respectively. Section \ref{concls} contains our conclusions.  The details of the calculation of the quadratic action are presented in the Appendix.

\section{Particles in theories with higher order derivatives}\label{preoncs}
The higher order derivative terms in the Lagrangian of a quantum field theory describe less local coupling at the level of the UV cutoff. This can be seen easily in lattice regularization where a finite difference of order $n$ in the action represents coupling of field variables, placed $n$ times the lattice spacing apart and renders the dynamics nonlocal up to these distances. Such a short distance modification of the dynamics used to arise in two different manners. One way is to implement a regulator which is kept as a physical law rather than being removed in the renormalization process. In fact, higher order derivatives in the kinetic energy have been used as possible regulators for a long time \cite{dirac,podolski,pauli,schwed,villars,feynman}. Another circumstance is when one is interested in the physics at energies well below the energy of the opening of particle creation channels and seeks an effective theory by eliminating the particle in question. The perturbative construction of such effective theories yields effective bare action with non-polynomial dependence in the space-time derivatives, a class of theories beyond our analytic capabilities. The truncation of the dependence on the space-time derivatives of the action into a polynomial of finite order renders the theory soluble within the scheme of perturbation expansion. 

The problem of theories with polynomial higher order derivatives, addressed in this Section, is a grave conflict with the probability interpretation of quantum mechanics.

\subsection{Difficulties}\label{diffs}
The source of the problems in theories with higher order derivatives is the unusually large set of frequencies which solve the linearized equation of motion. Let us consider this problem in the framework of a complex scalar field $\phi_j(x)$, $j=1,\ldots,n$ whose dynamics is defined by the Lagrangian
\be\label{befflbs}
L=\phid L(-\Box)\phi-V(\phid\phi),
\ee
characterized by two real functions, the local potential $V(\phid\phi)$ and $L(p^2)$. The interaction, $V''(\phid\phi)$, where the prime denotes the derivative with respect to the argument, $\phid\phi$, is supposed to be sufficiently weak to justify the discussion below and the kinetic energy, $L(p^2)$, is assumed to be a polynomial of order $n_d$. The mass term is placed into the potential by requiring $L(0)=0$. Let us assume a homogeneous external source, coupled linearly to the field which sets $\phi(x)=\bphi$. The dispersion relation of infinitesimal fluctuations around this state is given by the equation $\det[L(p^2)-{\cal M}^2]=0$ with
\be
{\cal M}^2=\openone V'(\bphid\bphi)+\bphi\otimes\bphid V''(\bar\phid\bphi).
\ee
Owing to the higher order derivative terms in the kinetic energy we have $2n_d$ different values of the frequency, corresponding to a fixed three-momentum $\ve{p}$. The generalization of usual construction of the Hamiltonian in classical mechanics \cite{ostrogadksi} leads to nondefinite kinetic energy in the Hamiltonian. 

The boundedness of the Hamiltonian from below can be saved in quantum mechanics by assuming negative norm in the subspace where the negative contributions of the kinetic energy are acting \cite{pais}. This can be seen easily in the case of the Lagrangian \eq{befflbs}, whose propagator,
\be
D=\frac1{L(p^2)-{\cal M}^2}
\ee
is given by the partial fraction decomposition as
\be\label{pvprop}
\frac1{L(p^2)-{\cal M}^2}=\sum_{n=1}^{n_d}\frac{z_n}{p^2-m_n^2},
\ee
in terms of the spectrum $\{m_n^2\}$, the set of the roots of $\det[L(p^2)-{\cal M}^2]$, and $z_n=[\partial L(p^2)/\partial p^2]^{-1}|_{p^2=m_n^2}$. It was assumed in deriving Eq.\eq{pvprop} that each eigenvalue $\lambda_a(p^2)$ of $L(p^2)-{\cal M}^2$ has single roots only. If the root $p^2=m_j^2$ of an eigenvalue is of $\ell$-th order then the right hand side may contain terms $z_{j,k}/(p^2-m_j^2)^k$ with $1\le k\le\ell$. It is attractive to interpret this form of the propagator by assuming that the field $\phi$ handles the propagation of several particles. Though the constants $z_n$ are real, they can be negative. In fact, any continuously derivable, real function, e.g. $x\to\lambda_a(x)$, displays alternating sign in its derivative as it crosses zero at odd order roots. If the negative contribution of ``particles'' with $z_n<0$ to the propagation is interpreted as the result of the negative norm of their state then their contribution to the kinetic energy becomes positive. Another problem, related to the unusual frequency spectrum is that poles $m_n^2$ may be complex and the unitarity of the time evolution is lost in this case.

One can argue that the negative norm states and the nonunitary amplitudes could be suppressed by a sufficiently low UV energy cutoff \cite{leewick}. But once negative norm states appear in the theory there are also states of vanishing norm and their virtual contribution to the transition amplitude diverges exponentially in time, free of any observable constraint. These modes can be eliminated by imposing appropriate boundary conditions at the final time \cite{leewick} but they in turn generate acausality and nonunitary time evolution, closing the vicious circle. The perturbative attempt to restore unitarity by modifying Feynman rules \cite{boulware} runs into difficulties, as well.

Negative norm states have already been used in gauge theories \cite{bleuer,gupta}, but these states can be excluded from the final state, as long as the system starts in the physically interpretable positive norm subspace, by means of gauge invariance. A similar symmetry argument was found useful to reach similar conclusion covering matter fields as well. One can show that the signature of norm of an excitation, created by the elementary physical fields is identical with product of the time and space reversal parities of the state \cite{cons,dyn}. Hence the unitarity of the time evolution follows within the positive norm subspace in time reversal invariant dynamics as long as the values $m_n^2$ are real. It is interesting to note that this argument is valid in quantum mechanics only. In classical physics a discrete symmetry leads to conservation law for harmonic systems only where the equation of motion is linear.

Higher order derivatives always occur in the effective bare action when the dynamics is sought below a particle creation threshold and their leading problems, mentioned above arise as soon as the effective action is truncated to polynomials of finite order in the gradient expansion. This is clearly a problem caused by the approximation only, the imposition of a UV cutoff below the heavy particle threshold solves all problems when the full effective action is used. But the systematic treatment of a sharp energy cutoff is rather difficult and one is usually satisfied by treating the higher order derivative terms as perturbation within the gradient expansion.

\subsection{Preons}\label{preonss}
The identification of a solution of a classical equation of motion, a differential equation of the order $2n_d$ in the time derivatives needs $2n_d$ parameters. This suggests that the field $\phi$ of the theory \eq{befflbs} describes $n_d$ particles, called pre-particles or preons below. The preon level description of this theory is defined by the partition function
\be
e^{iW[j,j^\dagger]}=\int\prod_nD[\phid_n]D[\phi_n]e^{i\int dx[\sum_n\phid_nD_n^{-1}\phi_n-V((\sum_n\phid_n)(\sum_n\phi_n))+j\sum_n\phid_n+j^\dagger\sum_n\phi_n]},
\ee
where it is the sum of the preon fields which enters into the vertices and which is coupled to the external source. This sum, denoted by $\phi$, can be introduced by a constraint, represented by the insertion of $1$ into the path integral,
\be
e^{iW[j,j^\dagger]}=\int D[\phid]D[\phi]\prod_nD[\phid_n]D[\phi_n]\prod_x\delta(\phi(x)-\sum_n\phi_n(x))e^{i\int dx[\sum_n\phid_nD_n^{-1}\phi_n-V(\phid\phi)+j\phid+j^\dagger\phi]}.
\ee
The Fourier integral representation of the Dirac-delta yields the form
\be
e^{iW[j,j^\dagger]}=\int D[\alpha]D[\alpha^\dagger]D[\phid]D[\phi]\prod_nD[\phid_n]D[\phi_n]e^{i\int dx[\sum_n\phid_nD_n^{-1}\phi_n-V(\phid\phi)+\alpha^\dagger(\phi-\sum_n\phi_n)+\alpha(\phid-\sum_n\phid_n)+j\phid+j^\dagger\phi]},
\ee
where the preon field can easily be integrated over,
\be
e^{iW[j,j^\dagger]}=\int D[\alpha]D[\alpha^\dagger]D[\phid]D[\phi]e^{i\int dx[-\alpha^\dagger\sum_nD_n\alpha-V(\phid\phi)+(\alpha^\dagger+j^\dagger)\phi+\phid(\alpha+j)]},
\ee
and one finds
\be
e^{iW[j,j^\dagger]}=\int D[\phid]D[\phi]e^{i\int dx[\phid D^{-1}\phi-V(\phid\phi)+j^\dagger\phi+\phid j]},
\ee
with
\be
D=\sum_nD_n,
\ee
after eliminating the auxiliary Fourier variables. 

The poles of the propagator in energy at vanishing momentum define the preon mass spectrum and we have $n_d$ preons if the function $L(p^2)$ has single roots only. The inequalities $m_n^2>0$ and $z_n>0$ are needed to guarantee unitary time evolution and positive definite norm for preons, respectively. In case of higher order roots the corresponding preon gives no contribution to scattering amplitude, it is ``confined''. 

The preons are not elementary constituents of the excitation of the field $\phi$, the latter can rather be imagined as appearing in the form of one of the preons, the relative weight is given by $z_n$. A similar situation can be found in QED where the one-loop photon self energy is resummed by solving the Schwinger-Dyson equation. The infinite geometrical series representation of the photon propagator can be interpreted as having infinite number of preons, corresponding to different number of particle-antiparticle pairs creation process.

\subsection{Preon condensate}\label{afvac}
In order to see better the low energy preon dynamics let us consider a theory whose effective action, the generator functional of one-particle irreducible graphs of the field $\phi$ is given by the space-time integral of the Lagrangian \eq{befflbs}-\eq{pvprop}. The infinitesimal fluctuations of $\phi(x)$ have relativistic dispersion relation with real mass $m_n$, satisfying the equation $\lambda_n(m_n^2)=0$. We restrict ourselves to the positive norm subspace where each preon has positive kinetic energy.

Let us change slowly, adiabatically the effective potential $V(\phid\phi)$ in such a manner that one or several $m_n^2$ approach zero and after that change sign. How does the vacuum change during this process? The particle number density, given by the propagator at zero separation in the leading order is the sum of the preon densities and the contribution of the light preons is obviously dominant when some of the preons become light. As soon as the preon mass squared becomes negative the vacuum energy is lowered by spontaneous preon-antipreon pairs creation until the repulsive interaction stabilizes the vacuum, a neutral coherent state. 

We look first for a static condensate and the possible time-dependent instabilities will be excluded by checking if small fluctuations have stable excitation spectrum next. The vacuum energy density is given by means of the expectation value of the energy-momentum tensor, $\la T^{00}\ra$ which is the sum of the tree-level and the loop contributions. The tree-level contribution, 
\be
\la T^{00}\ra_{tree}=\frac{\partial L}{\partial\partial_0\phi}\partial_0\phi+\partial_0\phid\frac{\partial L}{\partial\partial_0\phid}-L+{\cal E}_{ho}
\ee
contains a higher order derivative term, ${\cal E}_{ho}$. For static field configuration $\la T^{00}\ra_{tree}=-L$ because space derivatives are left in this case only and they contribute to the potential energy.

As long as preons with vanishing momentum build up the vacuum the condensate is space independent, homogeneous. When preon states with nonvanishing momentum are macroscopically occupied then the interactions of preons in the vacuum can be taken into account on the tree level, by minimizing the volume integral of the static energy density. This minimization is highly nontrivial due to the competition of the space-derivative terms, $\bphid L(\Delta)\bphi$ and the local potential, $V(\bphid\bphi)$. If $L(-\ve{p}^2)$ assumes its maximum at nonvanishing momentum then the configuration, minimizing the energy can be a space-dependent, periodic function. The simplest periodic vacuum configuration is a plane wave but oscillatory behavior can be observed in several directions, depending on the parameters of the theory according to the minimization of a 2-dimensional Euclidean field theory action, carried out in Ref. \cite{chekcer} which can be considered as the energy density of static configurations in 2+1 dimensional space-time.

\subsection{Goldstone preons}\label{godstprs}
It is natural to raise the question what happens when a continuous symmetry is broken spontaneously by the vacuum. Let us suppose for simplicity that the mass squared is non-negative for all preons and the vacuum is homogeneous and apply an infinitesimal symmetry transformation, $\bphi\to\bphi+\delta\bphi$, $\delta\bphi=i\epsilon\tau\bphi$, given in terms of a generator $\tau^\dagger=\tau$, which preserves the Lagrangian \eq{befflbs},
\be
V'(\bphi^\dagger\bphi)\bphi^\dagger\tau\bphi=0.
\ee
The derivative of this equation with respect to  $\bar\phi^\dagger$, evaluated in the vacuum,
\be
0={\cal M}^2\tau\bphi
\ee
is the usual Goldstone theorem for the mass spectrum. The dispersion relation of the Goldstone preons, 
\be
\det_{{\cal M}^2=0}[L(p^2)]=0
\ee
the vanishing of the determinant within the null-space of ${\cal M}^2$ contains a massless preon since $L(0)=0$. If $p^2=0$ is a single pole within the null-space then we have a massless preon but there might be additional massive Goldstone preons, as well. If $p^2=0$ is a higher order pole then the massless Goldstone preon is ``confined'', it does not correspond to a particle-like asymptotic state and scattering experiments see no massless Goldstone mode.

\section{Gauge model}\label{gaguges}
The use of higher order covariant derivatives in gauge theories opens up new possibilities concerning space-time symmetries, namely the recovery of translation invariance of a certain family of inhomogeneous vacua and the spontaneous breakdown of Lorentz symmetry \cite{dyn}. We shall discuss these issues in the framework of the model \eq{befflbs} by upgrading the global, semi-simple $U(n)$ symmetry group of a local $U(n)=SU(n)\times U(1)$ symmetry. The Lagrangian of this Yang-Mills-Higgs model is $L=L_m+L_g$, where
\bea\label{glagr}
L_g&=&-\hf\tr F^2-\frac14G^2,\nn
L_m&=&\phid L(-D^2)\phi-V(\phid\phi),
\eea
where the covariant derivative, $D_\mu=\partial_\mu-igA_\mu-ieB_\mu$, is given by means of the gauge fields $A_\mu=A_\mu^a\tau^a$ and $B_\mu$, with $[\tau^a,\tau^b]=if^{abc}\tau^c$ and $\tr\tau^a\tau^b=\hf\delta^{ab}$. The field strength tensors are defined as $F_{\mu\nu}=\partial_\mu A_\nu-\partial_\nu A_\mu-ig[A_\mu,A_\nu]=F^a_{\mu\nu}\tau^a$ and $G_{\mu\nu}=\partial_\mu B_\nu-\partial_\nu B_\mu$.

Let us suppose that the vacuum of the model with global symmetry, $g=e=0$ is of the form $\la\phi(x)\ra=\omega(x)\bphi_0$, where $\omega(x)$ is a local $U(n)$ transformation. The gauge transformation $\phi\to\omega^\dagger\phi$, $igA_\mu+ieB_\mu\to\omega(\partial_\mu-igA_\mu-ieB_\mu)\omega^\dagger$ brings this vacuum into the form $\la\phi(x)\ra=\bphi_0$, $\la igA_\mu(x)+ieB_\mu(x)\ra=\omega\partial_\mu\omega^\dagger$. A simple family of vacua which consists of a plane wave of an Abelian subgroup, $\omega(x)=e^{-ikx\tau}$, $i\tau$ being a generator of $U(n)$, yields $\la\phi(x)\ra=\bphi_0$ and $\la gA_\mu(x)+eB_\mu(x)\ra=k_\mu\tau$. The dynamical mechanism of generating nonvanishing gauge field from a scalar condensate in the vacuum is provided by the higher order derivative terms in the kinetic energy of $L_m$ which contains a local potential for the gauge field \cite{dyn} with degenerate minima. The latter makes the interaction between gauge bosons attractive in the naive, ``empty'' gauge vacuum and the gauge boson density is stabilized by their repulsive contact interaction. 

Once a particle with nonvanishing spin forms a homogeneous condensate the invariance with respect to the Lorentz group is broken. This leads to the loss of symmetry under spatial rotations or Lorentz boosts when the condensate is space- or time-like, respectively. We explore below vacua with unbroken spatial rotational symmetry because this symmetry is expected to be present at any scale and space-like condensate violates time reversal invariance \cite{cons}.

\subsection{Mean-field vacuum}\label{meanfvs}
The homogeneous, tree-level vacuum is based on the separation $\phi(x)=\bphi+\chi(x)$, $A_\mu(x)=A_\mu+a_\mu(x)$ and $B_\mu(x)=B_\mu+b_\mu(x)$  of the fields to a homogeneous condensate and quantum fluctuations. The basic building block of the kinetic energy of the scalar field, the argument of the function $L$ can be written as the sum of the mean-field and fluctuation contributions,
\be
D^2=\Box'-(a^a_\mu,b_\mu)\begin{pmatrix}g^2&eg\cr eg&e^2\end{pmatrix}\begin{pmatrix}a^{a\mu}\cr b^\mu\end{pmatrix}+a^a\alpha^a+b\beta
\ee
where $\Box'=\Box-(gA+eB)^2-2i(gA+eB)\partial$, 
\bea\label{abmfv}
\alpha^a&=&-g\tau^a(2i\partial+i\overleftarrow\partial+2eB)-g^2\{\tau^a,A\},\nn
\beta&=&-e(2i\partial+i\overleftarrow\partial+2eB+2gA).
\eea
The homogeneous, tree-level vacuum is found by minimizing the potential energy density
\be
U(A,B,\bar\phid,\bar\phi)=\hf\tr\left[(-ig[A_\mu,A_\nu])^2\right]-\bphid L((gA+eB)^2)\bphi+V(\bphid\bphi),
\ee
defined by omitting all space-time derivatives from the Lagrangian. We assume that the three-rotation remains a symmetry of the vacuum, hence the forms $A_\mu=n_\mu A$, $B_\mu=n_\mu B$, with $n^2=1$ follow, reducing the potential energy density to
\be\label{poten}
U(A,B,\bar\phid,\bar\phi)=V(\bar\phid\bar\phi)-\bar\phid L((gA+eB)^2)\bar\phi.
\ee

We shall consider the theory in static temporal gauge where the temporal components of the gauge field are static and behave as a three-dimensional adjoint scalar field with respect to the residual symmetry of this gauge, the space-dependent gauge transformations. Part of this symmetry is fixed by going into unitary gauge where these three-dimensional fields are diagonal, $(gA+eB)_{ij}=\delta_{ij}k_j$. The notation $k^\mu=n^\mu k$ will be used for the gauge field condensate, as well, $k=gA+eB$ being an $n\times n$ matrix. The vacuum energy density can now be recast in the form
\be
U(A,B,\bphid,\bphi)=V(\bphid\bphi)-\sum_{j=1}^n\bphi_j^\dagger L(k_j^{2})\bphi_j.
\ee
The non-diagonal components of the gauge field are charged with respect to the remaining Abelian $U(1)^n$ gauge symmetry which is further restricted by taking the scalar condensate, $\bphi$, real.

We are interested in the case $gA+eB\ne0$ therefore we need $\bphi\ne0$. The gauge and the scalar field condensates may belong to different subspaces of the symmetry $U(n)$. If $m=\mr{Rang}(gA+eB)<n$ then we have the standard Higgs mechanism for $n-m$ components of the Higgs field. We look for the interplay of the scalar and the gauge condensate therefore $\mr{Rang}(gA+eB)=n$ will be assumed in what follows. Minimization with respect to the gauge field gives 
\be\label{gmin}
L'(k_j^{2})=0
\ee
and $L''(k_j^{2})<0$ as long as $\bphi_j\ne0$. The minimization in the scalar field yields the condition
\be\label{smin}
L(k^2)=V'(\bphid\bphi).
\ee
Therefore, all $k_j$ components corresponding to $\bphi_j\ne0$ must be at the maximum of $L(p^2)$. The tree-level vacuum allows global gauge rotations as residual symmetries within the degenerate subspaces of $k_j$. Different distribution of the magnitude of the scalar condensate among these degenerate subspaces yields inequivalent, classically degenerate vacua.

\subsection{Fluctuations}\label{flucts}
The determination of the spectrum of infinitesimal fluctuations serves two goals, it establishes the stability of the static vacuum with respect time dependent perturbation and gives the particle content of the theory. 

It is advantageous to separate the real and imaginary part of the scalar field, $\chi=\chi_1+i\chi_2$ and use the Fourier representation
\be
f(x)=\int_pf(p)e^{-ipx},
\ee
with $\int_p=\int d^4p/(2\pi)^4$ for the fields. The calculation, sketched in Appendix \ref{flucta} gives the quadratic action
\bea\label{quadract}
S^{(2)}&=&\int_p(\chi_1(-p),\chi_2(-p),a^a_\mu(-p),b_\mu(-p))\nn
&&\begin{pmatrix}\begin{pmatrix}L^+_d&iL^-_d\cr -iL^-_d& L^+_d\end{pmatrix}-2V''\begin{pmatrix}\bphi_1\otimes\bphi_1&0\cr0&0\end{pmatrix}&
\begin{pmatrix}[L_dM^{b\nu}]&[L_dN^\nu]^+\cr [iL_dM^{b\nu}]&[iL_dN^\nu]^-\end{pmatrix}\bphi_1\cr
\bphid_1\begin{pmatrix}[M^{a\mu}L_d]&[-iM^{a\mu}L_d]\cr[N^\mu L_d]^+&-i[N^\mu L_d]^-\end{pmatrix}&G_g+G_m\end{pmatrix}
\begin{pmatrix}\chi_1(p)\cr\chi_2(p)\cr a^b_\nu(p)\cr b_\nu(p)\end{pmatrix}
\eea
where $G_g$ and $G_m$ are given by Eqs. \eq{abquadr} and \eq{abquadrm}, respectively, $L_d=L((p+k)^2)-L(k^2)$ and
\bea
M^{a\mu}&=&\frac{g(\tau^ap^\mu+n^\mu\{\tau^a,k\})}{p^2+2pnk},\nn
N^\mu&=&\frac{e(p^\mu+2kn^\mu)}{p^2+2pnk}.
\eea
The use of real field components in space-time leads to the combinations $f(p)^+=\hf[f(p)+f(-p)]$ for real and $f(p)^-=\hf[f(p)-f(-p)]$ for imaginary matrix elements.

We impose now static temporal gauge by writing $a^\mu=(a^0,\ve{n}a_L+\ve{a}_T)$, $b^\mu=(b^0,\ve{n}b_L+\ve{b}_T)$, $\ve{n}=\ve{p}/|\ve{p}|$ in Fourier space. The quadratic action assumes the form
\bea\label{quadractstg}
S^{(2)}&=&\int_p(\chi_1(-p),\chi_2(-p),a^a_L(-p),b_L(-p))K(p)
\begin{pmatrix}\chi_1(p)\cr\chi_2(p)\cr a^b_L(p)\cr b_L(p)\end{pmatrix}
+\hf\int_p(\ve{a}^a_T(-p),\ve{b}_T(-p))G_T(p)\begin{pmatrix}\ve{a}^b_T(p)\cr\ve{b}_T(p)\end{pmatrix}+\tilde{S}^{(2)}
\eea
where
\be
K(p)=\begin{pmatrix}\begin{pmatrix}L^+_d&iL^-_d\cr -iL^-_d& L^+_d\end{pmatrix}-2V''\begin{pmatrix}\bphi_1\otimes\bphi_1&0\cr0&0\end{pmatrix}&
\begin{pmatrix}-g[L_dN\tau^b]^\pm&-e[L_dN]^+\cr -ig[L_dN\tau^b]^\pm&-ie[L_dN]^-\end{pmatrix}\bphi_1\cr
\bphid_1\begin{pmatrix}-g[\tau^aNL_d]^\pm&ig[\tau^aNL_d]^\pm\cr-e[NL_d]^+&ie[NL_d]^-\end{pmatrix}&G^\pm_g+G^\pm_m\end{pmatrix},
\ee
$N=|\ve{p}|/(p^2+2pnk)$ and $\tilde{S}^{(2)}$ denotes the contribution of the static modes, an expression similar to \eq{quadract} except that the integration is over three momentum and the fields and the quadratic form is taken at vanishing frequency. The subscript $\pm$  of the complex matrix elements, acting on the non-Abelian gauge field is a reminder that the real and the imaginary components should be symmetrized or antisymmetrized, respectively. An important simplification comes from the fact that the higher order derivative terms in the kinetic energy, being constructed by means of the four-vectors $p^\mu$ and $n^\mu$ do not act on the transverse gauge field components whose action comes entirely form $L_g$,
\be\label{trquform}
G^{\mu\nu}_T=\left(g^{\mu\nu}-\frac{p^\mu p^\nu}{p^2}\right)\begin{pmatrix}-\delta^{ab}p^2+g^2f^{dca}f^{deb}A^cA^e+2igf^{abc}A^cnp&0\cr0&-p^2\end{pmatrix},
\ee
according to Eq. \eq{abquadr}. 

The most important information in the quadratic action is the preon dispersion relation, given by the set of four vectors $p^\mu=(\omega,\ve{p})$ where some matrix elements of the propagators $K^{-1}(p)$ or $G_T^{-1}(p)$ diverge. The inverse matrix element of a matrix is the ratio of a minor and the determinant. The former is a polynomial of finite order in $p$, hence the propagator can display singularity at the root of the fluctuation determinant, $\det K(p)$ or $\det G_T(p)$.

The spectrum of the transverse gauge field components can easily be determined by noting that the action for these components comes from $G_g$ alone which is free of higher order derivatives. Therefore the gauge field condensate acts as a chemical potential and shifts the preon energies of the charged, non-diagonal gauge fields. It is advantageous to use the charge-diagonal basis, defined by $[A,\tau^a]=q^a\tau^a$ because the quadratic form \eq{trquform} is diagonal and the diagonal elements, $(\omega-gq^a)^2-\ve{p}^2$, give the massless spectrum, shifted by a chemical potential, $\omega=|\ve{p}|+gq^a$. 

It is instructive to compare the spectrum with those, predicted with the conventional Higgs mechanism. The most obvious difference is that the transverse gauge field components remain massless, the term which is responsible for the gauge boson mass generation in the usual Higgs mechanism, \eq{uhmech}, is absent from \eq{quadract} due to  Eq. \eq{gmin}, the possibility of accumulating gauge bosons in macroscopic quantity in the vacuum. Another remarkable feature of this generalized Higgs mechanism is that the preon spectrum breaks Lorentz invariance by a shift of the energy, a chemical potential only.

\subsection{Gapless scalar preon spectrum}\label{gaplesss}
As the simplest use of the action \eq{quadractstg} we determine the gap in the preon spectrum. The gap is the energy at vanishing momentum hence we need the quadratic action for homogeneous, time dependent fields only. The static modes represent no dynamical degrees of freedom and $\tilde{S}^{(2)}$ can be neglected in \eq{quadractstg}, leaving the quadratic action
\bea
S^{(2)}&=&\int\frac{d\omega}{2\pi}(\chi_1(-\omega),\chi_2(-\omega))\begin{pmatrix}L^+_d-2V''\bphi_1\otimes\bphi_1&iL^-_d\cr -iL^-_d& L^+_d\end{pmatrix}\begin{pmatrix}\chi_1(\omega)\cr\chi_2(\omega)\end{pmatrix}\nn
&&+\hf\int\frac{d\omega}{2\pi}(a^a_L(-\omega),b_L(-\omega))\left[\omega^2-\begin{pmatrix}g^2f^{dca}f^{deb}A^cA^e+2igf^{abc}A^c\omega&0\cr0&0\end{pmatrix}\right]\begin{pmatrix}a^b_L(\omega)\cr b_L(\omega)\end{pmatrix}
\eea
where the scalar and gauge fields do not mix. 

The action of the longitudinal gauge fields comes from $G_g$ and the remark, mentioned above about the transverse spectrum applies, giving the spectrum $(\omega-gq^a)^2$, displaying nonvanishing gap, $\pm gq^a$, for the non-diagonal, charged components. The scalar sector is gapless, as expected by Goldstone theorem, owing to the vanishing of the fluctuation determinant for $\omega=\ve{p}=0$,
\be
\det\left|\begin{matrix}-2V''\bphi_1\otimes\bphi_1&0\cr0&0\end{matrix}\right|=0
\ee
which is the product of the scalar preon masses.

\section{$U(2)$ gauge theory}\label{utgths}
It is instructive to consider the simplest nontrivial realization of the model discussed above, using the gauge group $U(2)$, in more details and check the conditions of unitarity. We take the simplest choice of the scalar kinetic energy with a maximum,
\be
L(p^2)=L_0-\frac{(p^2-\kappa^2)^2}{\kappa^2}
\ee
giving $L_d(p)=-((p+\kappa)^2-\kappa^2)^2/\kappa^2$ and $k_j=\pm \kappa$. In case of identical signs, $k_j=(\pm \kappa,\pm \kappa)$ we have Abelian condensate, $eB=\pm \kappa$. When $k_j=(\pm \kappa,\mp \kappa)$ then the gauge condensate is non-Abelian, $gA^a=2\kappa\delta^{a3}$. There is no place for condensate containing both Abelian and non-Abelian components with such a simple kinetic energy and there are two, tree-level degenerate, inequivalent families of vacua.

\subsection{Non-Abelian condensate}\label{nagcs}
Let us assume that the gauge field condensate is non-Abelian, $B_\mu=0$ and
\be
gA^\mu=n^\mu\begin{pmatrix}1&0\cr 0&-1\end{pmatrix}\kappa,
\ee
satisfying Eq. \eq{gmin}. The scalar condensate $(\bphi_1,\bphi_2)=(\sin\theta,\cos\theta)\bphi$, where real $\bphi$ is chosen in such a manner that Eq. \eq{smin} is satisfied. The vacua corresponding to different values of $\theta$ are inequivalent and degenerate on the tree-level.

The first term of the quadratic action, \eq{quadractstg}, can be written as
\be\label{qaac}
S^{(2)}=\hf\int_p(h(-p),g(-p))\begin{pmatrix}K_{hh}(p)&K_{hg}(p)\cr K_{gh}(p)&K_{gg}(p)\end{pmatrix}\begin{pmatrix}h(p)\cr g(p)\end{pmatrix},
\ee
where $h=(\Re\chi^1,\Im\chi^1,\Re\chi^2,\Im\chi^2)$, $g=(a^1_L,a^2_L,a^3_L,b_L)$ and
\bea\label{kgh}
K_{hh}&=&\begin{pmatrix}2L_d^+-4V''\bphi^2\sin^2\theta&2iL_d^-&-2V''\bphi^2\sin2\theta&0\cr
-2iL_d^-&2L_d^+&0&0\cr
-2V''\bphi^2\sin2\theta&0&2L_d^+-4V''\bphi^2\cos^2\theta&-2iL_d^-\cr
0&0&2iL_d^-&2L_d^+\end{pmatrix},\nn
K_{hg}&=&\begin{pmatrix}-gR^-\cos\theta&igR^+\cos\theta&-gR^-\sin\theta&-2eR^-\sin\theta\cr
igR^+\cos\theta&gR^-\cos\theta&igR^+\sin\theta&2ieR^+\sin\theta\cr
gR^-\sin\theta&-igR^+\sin\theta&-gR^-\cos\theta&2eR^-\cos\theta\cr
igR^+\sin\theta&gR^-\sin\theta&-igR^+\cos\theta&2ieR^+\cos\theta\end{pmatrix}=K_{gh}^\dagger,\nn
K_{gg}&=&\begin{pmatrix}
\omega^2+4\kappa^2+g^2S^+&-4i\omega \kappa+ig^2S^-\cos 2\theta&0&2egS^+\sin2\theta&\cr
4i\omega \kappa-ig^2S^-\cos 2\theta&\omega^2+4\kappa^2+g^2S^+&0&2iegS^-\sin2\theta&\cr
0&0&\omega^2+g^2S^+&-2egS^+\cos2\theta&\cr 2egS^+\sin2\theta&-2iegS^-\sin2\theta&-2egS^+\cos2\theta&\omega^2+4e^2S^+\cr\end{pmatrix},
\eea
with 
\bea
R^\pm&=&\frac{\bphi|\ve{p}|}2\left[\frac{L_d(p)}{(p^2+2\omega \kappa)}\pm\frac{L_d(-p)}{(p^2-2\omega \kappa)}\right]\ ,\nn
S^\pm&=&\frac{\bphi^2|\ve{p}|^2}4\left[\frac{L_d(p)}{(p^2+2\omega \kappa)^2}\pm\frac{L_d(-p)}{(p^2-2\omega \kappa)^2}\right]\ .
\eea

The theory has unitary time evolution if all preon masses are real. The determinant of the quadratic form in Eq. \eq{qaac},
\bea
\det K&=&-\frac{16}{\kappa^8}\omega^4(\omega-2\kappa)^2(\omega+2\kappa)^2(\omega^2-2\omega \kappa-\ve{p}^2)^2(\omega^2+2\omega \kappa-\ve{p}^2)^2\nn
&&\times[2\kappa^2\bphi^2V''(2g^2\ve{p}^2\bphi^2-\omega^4+2\ve{p}^2\omega^2-4\kappa^2\omega^2-(\ve{p}^2)^2+8e^2\ve{p}^2\bphi^2\cos^22\theta)\nn
&&-\omega^8+4\ve{p}^2\omega^6+8\kappa^2\omega^6-6(\ve{p}^2)^2\omega^4-16\kappa^2\ve{p}^2\omega^4-16\kappa^4\omega^4+4(\ve{p}^2)^3\omega^2+8\kappa^2(\ve{p}^2)^2\omega^2-(\ve{p}^2)^4]
\eea
has complex frequency roots unfortunately. The source of trouble, the dependence on the scalar condensate, can be eliminated by imposing a further condition \cite{dyn},
\be\label{ftune}
V''=0,
\ee
when one finds
\be
\det K=\frac{16}{\kappa^8}\omega^4(\omega-2\kappa)^2(\omega+2\kappa)^2\left(\omega^2-2\omega \kappa-\ve{p}^2\right)^4\left(\omega^2+2\omega \kappa-\ve{p}^2\right)^4,
\ee
and the preon spectrum is real and gapless, $\omega_{\sigma,\sigma'}=\sigma \kappa+\sigma'\sqrt{\kappa^2+\ve{p}^2}$, with $\sigma,\sigma'=\pm1$. 

Let us now turn to the gauge boson propagation. The form \eq{trquform} gives a massless transverse propagator with shifted energies for the charged gauge field components,
\be
G^{-1\mu\nu}_T=\left(g^{\mu\nu}-\frac{p^\mu p^\nu}{p^2}\right)\begin{pmatrix}\delta^{ab}\frac1{(\omega-gq^a)^2-\ve{p}^2}&0\cr0&\frac1{p^2}\end{pmatrix}\ .
\ee
written in the charge diagonal basis, $a^a=(a^+,a^-,a^3)$, with $a^\pm=(a^1\mp ia^2)/\sqrt2$ and $q^a=(2\kappa/g,-2\kappa/g,0)$. To find the longitudinal propagator we first eliminate the scalar field from the quadratic action by means of its equation of motion. The elimination of $\chi^1$ and $\chi^2$ generates the term
\be
\hf\int_p(a_L^+(-p),a_L^-(-p))\begin{pmatrix}(\omega-2\kappa)^2&0\cr 0&(\omega+2\kappa)^2\end{pmatrix}\begin{pmatrix}a_L^-(p)\cr a_L^+(p)\end{pmatrix},
\ee
with shifted energies for the charged fields. 
It also yields the contribution
\be\label{ablqa}
\hf\int_p\omega^2[b_L(-p)b_L(p)+a^3_L(-p)a^3_L(p)]\ ,
\ee
to the action. The quadratic action for $a^3_L$ and $b_L$
is diagonal when the field combinations 
\bea\label{diagab}
a_L&=&\frac{g'b_L+e'a^3_L}{\sqrt{g'^2+e'^2}},\nn
z_L&=&\frac{-e'b_L+g'a^3_L}{\sqrt{g'^2+e'^2}},
\eea
with $g'=g\cos2\theta$, $e'=2e$ are used,
\be\label{ablqaf}
\hf\int_p\omega^2[a_L(-p)a_L(p)+z_L(-p)z_L(p)].
\ee

The theory, obtained by the fine tuning \eq{ftune} is unitary in the positive norm, physical sector and has two charged scalar preons for each scalar field component, $\phi_j$. At least one of these field components has gapless excitation spectrum according to Goldstone theorem. There are two, propagating massless transverse helicity preon modes for each gauge field $a^a_\mu$ and $b_\mu$. The longitudinal gauge field components have a single, non-propagating and gapless preon mode as expected from the space-dependent residual symmetry of static temporal gauge. The energy spectrum of the charged components of the non-Abelian gauge field is shifted by the gauge field condensate as chemical potential. The absence of the parameter $\theta$ of the vacuum in the eigenvalues of the propagator indicates that the vacua with different values of $\theta$ remain degenerate on the one-loop level.

\subsection{Abelian condensate}\label{agcs}
Another family of vacua is where the Abelian gauge field develops vacuum expectation value, $A_\mu=0$ and $eB^\mu=n^\mu \kappa$. The residual global $U(2)$ can be used to rotate the scalar condensate into the form $\bphi_j=(0,\bphi)$, with real $\bphi$. The action, \eq{qaac} now contains the matrices:

\bea\label{kghh}
K_{hh}&=&\begin{pmatrix}2L_d^+&2iL_d^-&0&0\cr
-2iL_d^-&2L_d^+&0&0\cr
0&0&2L_d^+-4V''(\bphi)^2&2iL_d^-\cr
0&0&-2iL_d^-&2L_d^+\end{pmatrix},\nn
K_{hg}&=&\begin{pmatrix}-gR^-&igR^+&0&0\cr
igR^+&gR^-&0&0\cr
0&0&gR^-&-2eR^-&\cr
0&0&-igR^+&2ieR^+&\end{pmatrix}=K_{gh}^\dagger,\nn
K_{gg}&=&\begin{pmatrix}
\omega^2+g^2S^+&-ig^2S^-&0&0\cr
ig^2S^-&\omega^2+g^2S^+&0&0\cr
0&0&\omega^2+g^2S^+&-2egS^+\cr
0&0&-2egS^+&\omega^2+4e^2S^+\end{pmatrix}\ .
\eea
The fluctuation determinant reads
\bea
\det K&=&-\frac{16}{\kappa^8}\omega^8\left(\omega^2-2\omega \kappa-|\ve{p}|^2\right)^2\left(\omega^2+2\omega \kappa-|\ve{p}|^2\right)^2[2\kappa^2(\bphi)^2V''(2g^2\ve{p}^2(\bphi)^2+8e^2\ve{p}^2(\bphi)^2-\omega^4+2\ve{p}^2\omega^2-4\kappa^2\omega^2\nn
&&-(\ve{p}^2)^2)-\omega^8+4\ve{p}^2\omega^6+8\kappa^2\omega^6-6(\ve{p}^2)^2\omega^4-16\kappa^2\ve{p}^2\omega^4-16\kappa^4\omega^4+4(\ve{p}^2)^3\omega^2+8\kappa^2(\ve{p}^2)^2\omega^2-(\ve{p}^2)^4].
\eea
It gives complex frequencies as in the case of non-Abelian condensate and we impose again the fine tuning condition \eq{ftune}, yielding
\be
\det K=\frac{\omega^8}{\kappa^8}(\omega^2-2\omega\kappa-\ve{p}^2)^4(\omega^2+2\omega \kappa-\ve{p}^2)^4,
\ee
to establish unitary time evolution. The preon spectrum is the same as in the case of non-Abelian condensate, namely $\omega_{\sigma,\sigma'}=\sigma \kappa+\sigma'\sqrt{\kappa^2+\ve{p}^2}$. Any other $L(p^2)=\ord{p^4}$ polynomial which is bounded from above and has maximum at $\kappa^2$ reproduces the same fluctuation determinant up to a constant factor.

The transverse gauge boson propagator, \eq{trquform} is 
\be
G^{-1\mu\nu}_T=\frac1{p^2}\left(g^{\mu\nu}-\frac{p^\mu p^\nu}{p^2}\right)\begin{pmatrix}\delta^{ab}&0\cr0&1\end{pmatrix}\ .
\ee
The elimination of $\chi^1$ from the quadratic action by means of its equation of motion generates a term
\be
\int_p\omega^2a^+_L(-p)a^-_L(p).
\ee
When $\chi^2$ is eliminated then the contribution \eq{ablqa} is generated in the action. The full quadratic action for $a^3_L$ and $b_L$ can be diagonalized by the help of the combinations \eq{diagab} with $\theta=0$ and \eq{ablqaf} is recovered. The spectrum is similar as in the vacuum with non-Abelian condensate except that the chemical potential which shifted there the energies is vanishing in this case.

\section{Conclusions}\label{concls}
Models with higher order derivatives are usually considered as effective theories, arising by the elimination of some heavy particles. A different motivation to consider such theories is followed in this work, namely the higher order derivative terms not only preserve renormalizability but act as a physical UV cutoff and opens the way to apply a wide family of new theories to realistic problems. 

The case of a spontaneously broken global gauge invariance is considered in this work where the higher order covariant derivatives in the scalar field kinetic energy generate a local potential for the gauge field which in turn induces a nonvanishing expectation value of the gauge field. In order to preserve rotational invariance and unitarity only the temporal component of the gauge field can develop expectation value.

In contrast to the usual Higgs mechanism the gauge field remains massless and the gauge field condensate acts as a chemical potential only. The mass term is removed by the stabilization of the vacuum with respect to the density of gauge bosons in the coherent state vacuum. In other words, the spontaneous breakdown of the symmetry with respect to Lorentz boosts consists of the identification of a preferred inertial system where preons have relativistic dispersion relation with a chemical potential and form a relativistic gas at finite density.

  The higher order derivative terms in the kinetic energy suppress the UV dynamics and generate several dispersion relations, preons, for a single field. Therefore these terms give a new class of unified models. For instance one may reopen the issue of massive gauge bosons without spontaneous symmetry breaking, Higgs mechanism and renormalizable models can be constructed where the price of massive vector bosons is the appearance of further preons, particle-like modes in the theory.

\appendix
\section{Quadratic action}\label{flucta}
The quadratic part of the action, defined by the Lagrangians \eq{glagr} is  calculated in this Appendix, following the scheme worked out for the Abelian case \cite{dyn}.

\subsection{$L_g$}
Let us start with
\be
S_g=-\hf\int dx\tr[(\partial_\mu(A_\nu+a_\nu)-\partial_\nu(A_\mu+a_\mu)-ig[A_\mu+a_\mu,A_\nu+a_\nu])^2]-\frac14\int dx(\partial_\mu b_\nu-\partial_\nu b_\mu)^2
\ee
which gives the quadratic part
\bea
S^{(2)}_g&=&\int dx\biggl[-\hf\tr[(\partial_\mu a_\nu-\partial_\nu a_\mu)^2]+ig\tr[[a_\mu,a_\nu]F_{\mu\nu}[A]]+2ig\tr[(\partial_\mu a_\nu-\partial_\nu a_\mu)[A_\mu,a_\nu]]\nn
&&+\frac{g^2}2\tr[([A_\mu,a_\nu]+[a_\mu,A_\nu])^2]-\frac14(\partial_\mu b_\nu-\partial_\nu b_\mu)^2\biggr],
\eea
which can be written for the gauge field, given by the Fourier integral as
\be
S^{(2)}_g=\int_p(a_\mu(-p),b_\mu(-p))G^{\mu\nu}_g\begin{pmatrix}a_\nu(p)\cr b_\nu(p)\end{pmatrix}
\ee
where
\be\label{abquadr}
G^{\mu\nu}_g=\hf\openone(p^\mu p^\nu-p^2g^{\mu\nu})+\begin{pmatrix}\tilde G^{\mu\nu}_g&0\cr0&0\end{pmatrix}
\ee
and
\be
\tilde G^{(a\mu)(b\nu)}_g=\frac{g^2}2f^{dca}f^{deb}A^cA^e(n^\mu n^\nu+g^{\mu\nu})+\frac{ig}2f^{abc}A^c[2(n\cdot p)g^{\mu\nu}-n^\mu p^\nu-n^\nu p^\mu].
\ee

\subsection{$L_m$}
The scalar field kinetic energy contributes to the quadratic action by a number of terms which will be calculated by using the power series representation
\be
L(z)=\sum_{n=0}^\infty c_nz^n.
\ee

The quadratic part in the gauge fields comes from
\be\label{quadrgf}
\bphid L(-D^2)\bphi=\sum_nc_n\bphid[-(\partial-ieA-ieB)^2]^n\bphi.
\ee
One kind of contribution is found by keeping the quadratic part of $D^2$ in the gauge fields,
\be
\int_p\sum_nc_n\sum_{\ell=1}^n\bphid(-\Box')\cdots(a(-p),b(-p))\begin{pmatrix}g^2&eg\cr eg&e^2\end{pmatrix}\begin{pmatrix}a(p)\cr b(p)\end{pmatrix}\cdots(-\Box')\bphi.
\ee
which can be written as
\be\label{uhmech}
\int_p\bphid\sum_nc_nn(a(-p),b(-p))\begin{pmatrix}g^2&eg\cr eg&e^2\end{pmatrix}\begin{pmatrix}a(p)\cr b(p)\end{pmatrix}k^{2(n-1)}\bphi
=\bphid L'(k^2)\bphi\int_p(a(-p),b(-p))\begin{pmatrix}g^2&eg\cr eg&e^2\end{pmatrix}\begin{pmatrix}a(p)\cr b(p)\end{pmatrix}.
\ee
This term is vanishing due to Eq. \eq{gmin}.

The other quadratic contribution of \eq{quadrgf} arises by keeping the linear part from two different covariant derivatives,
\be
\int_p\sum_nc_n\sum_{\ell=1}^{n-1}\sum_{\ell'=\ell+1}^n\bphid k^2\cdots[a^a(-p)\alpha^a(-p)+b(-p)\beta(-p)]\cdots(p+k)^2\cdots[a^a(p)\alpha^a(p)+b(p)\beta(p)]\cdots k^2\bphi,
\ee
where $\alpha^a(p)=-g(\tau^ap+\{\tau^a,k\})$ and $\beta=-e(p+2k)$, cf. Eq. \eq{abmfv}. We rewrite this expression in a simpler manner by means of the functions $X(p)=a^a(p)\alpha^a(p)+b(p)\beta(p)$ and $f_{jm\ell}=f(k_j^2,(p+k_m)^2,k_\ell^2)$ as,
\be
\sum_{j\ell m}f_{jm\ell}\bphid_jX_{-pjm}X_{pm\ell}\bphi_\ell
\ee
where
\be
f(a,b,c)=\sum_nc_n\sum_{\ell=1}^{n-1}\sum_{\ell'=\ell+1}^na^{\ell-1}b^{\ell'-\ell-1}c^{n-\ell'}.
\ee
The sums can easily be calculated with the result
\be
f(a,b,c)=\frac{L(c)(b-a)+L(a)(c-b)-L(b)(c-a)}{(c-b)(c-a)(b-a)}
\ee
for $a\ne c$ which in the limit $a,c\to k^2_j$ becomes
\be
f(k_j,b,k_j)=\frac{L(b)-L(k^2_j)}{(b-k^2_j)^2}
\ee
by the help of Eq. \eq{gmin}. The $\ord{(a+b)^2}$ contribution is therefore
\be
\int_p(a_{-p}^a,b_{-p})G_m\begin{pmatrix}a_p^b\cr b_p\end{pmatrix},
\ee
with
\be\label{abquadrm}
G^{\mu\nu}_m=\sum_{j\ell m}f(k_j^2,(p+k_m)^2,k_\ell^2)\bphid_j\bphi_\ell\begin{pmatrix}\tilde G^{(a\mu)(b\nu)}_{j\ell m}&\tilde G^{(a\mu)\nu}_{j\ell m}\cr\tilde G^{\mu(b\nu)}_{j\ell m}&\tilde G^{\mu\nu}_{j\ell m}\end{pmatrix}
\ee
and
\bea
\tilde G^{(a\mu)(b\nu)}_{j\ell m}&=&g^2(\tau^ap^\mu+n^\mu\{\tau^a,k\})_{jm}(\tau^bp^\nu+n^\nu\{\tau^a,k\})_{m\ell}\nn
\tilde G^{(a\mu)\nu}_{j\ell m}&=&ge(\tau^ap^\mu+n^\mu\{\tau^a,k\})_{jm}(p^\nu+2kn^\nu)_{m\ell}\nn
\tilde G^{\mu(b\nu)}_{j\ell m}&=&eg(p^\mu+2kn^\nu)_{jm}(\tau^bp^\nu+n^\nu\{\tau^a,k\})_{m\ell}\nn
\tilde G^{\mu\nu}_{j\ell m}&=&e^2(p^\mu+2kn^\mu)_{jm}(p^\nu+2kn^\nu)_{m\ell}.
\eea

The mixing between the scalar and the gauge field comes from
\be
L(-D^2)\bphi=\sum_nc_n[-(\partial-igA-igB)^2]^n\bphi
\ee
whose linear part in the gauge field is
\be
-\sum_nc_n\sum_{\ell=1}^n(-\Box')\cdots(a^a\alpha^a+b\beta)\cdots(-\Box')\bphi.
\ee
hence the $\ord{\chid(a+b)}$ contribution of the scalar field kinetic term is
\be\label{ochidab}
-\chid\sum_nc_n\sum_{\ell=1}^n(p+k)^{\ell-1}X_pk^{2(n-\ell)}\bphi
=-\sum_{jm}\chid_jC((p+k_j)^2,k_m^2)X_{pjm}\bphi_m.
\ee
The summation over $\ell$ is trivial and we find
\bea
C(a,b)&=&\sum_nc_n\sum_{\ell=1}^na^{\ell-1}b^{n-\ell}\nn
&=&\frac{L(a)-L(b)}{a-b},
\eea
yielding
\be
-\chid[L((p+k)^2)Y_p-Y_pL(k^2)]\bphi
\ee
with $Y_{pjm}=X_{pjm}/[(p+k_j)^2-k_m^{2}]$ for \eq{ochidab}. The contribution $\ord{(a+b)\chi}$, 
\be
-\bphid[Y_{-p}L((p+k)^2)-L(k^2)Y_{-p}]\chi.
\ee
can be found in a similar manner.

Finally, the kinetic energy for the scalar field fluctuation assumes the form
\be
\int_p\left[\chid(-p)L_d(p)\chi(p)-\hf[\bphid\chi(-p)+\chid(-p)\bphi][\bphid\chi(p)+\chid(p)\bphi]V''(\bphid\bphi)\right]
\ee
with $L_d(p)=L((p+k)^2)-L(k^2)$, after using Eq. \eq{smin}.

\end{document}